\documentclass[aps,prb,preprint,longbibliography,superscriptaddress,11pt]{revtex4-2}
\usepackage{color}
\usepackage{xcolor}
\usepackage{graphicx}
\usepackage{amsmath}
\usepackage{amsfonts}
\usepackage{soul}
\usepackage[normalem]{ulem}
\usepackage{lipsum}
\usepackage{bibentry}
\usepackage{float}
\usepackage{silence}
\nobibliography*

\begin{document}

\title{Bose-Einstein condensation of an optical thermodynamic system into a solitonic state}

\author{Jiaxuan Zhang}%
\thanks{These two authors contributed equally}
\author{Jintao Fan}%
\thanks{These two authors contributed equally}
\affiliation{Ultrafast Laser Laboratory, Key Laboratory of Opto-electronic Information Science and Technology of Ministry of Education, School of Precision Instruments and Opto-electronics Engineering, Tianjin University, 300072 Tianjin, China}%
\author{Chao Mei}
\affiliation{Department of Physics, School of Physical Science and Technology, Ningbo University, 315211, Zhejiang, China}%
\author{G{\"u}nter Steinmeyer}
\email{steinmey@mbi-berlin.de}
\affiliation{Max Born Institute for Nonlinear Optics and Short Pulse Spectroscopy, 12489 Berlin, Germany}%
\affiliation{Institut f{\"u}r Physik, Humboldt Universität zu Berlin, 12489 Berlin, Germany}%
\author{Minglie Hu}
\email{huminglie@tju.edu.cn}
\affiliation{Ultrafast Laser Laboratory, Key Laboratory of Opto-electronic Information Science and Technology of Ministry of Education, School of Precision Instruments and Opto-electronics Engineering, Tianjin University, 300072 Tianjin, China}%
\date{\today}

\begin{abstract}
Recent years have seen a resurgence of interest in multimode fibers due to their intriguing physics and applications, with spatial beam self-cleaning (BSC) having received special attention. In BSC light condenses into the fundamental fiber mode at elevated intensities. Despite extensive efforts utilizing optical thermodynamics to explain such a counterintuitive beam reshaping process, several challenges still remain in fully understanding underlying physics. Here we provide compelling experimental evidence that BSC in a dissipative dual-core fiber can be understood in full analogy to Bose-Einstein condensation (BEC) in dilute gases. Being ruled by the identical Gross-Pitaevskii Equation, both systems feature a Townes soliton solution, for which we find further evidence by modal decomposition of our experimental data. Specifically, we observe that efficient BSC only sets in after an initial thermalization phase, causing convergence towards a Townes beam profile once a threshold intensity has been surpassed. This process is akin to a transition from classical to quantum-mechanical thermodynamics in BEC. Furthermore, our analysis also identifies dissipative processes as a crucial, yet previously unidentified component for efficient BSC in multimode fiber. This discovery paves the way for unprecedented applications of multimode-fiber based systems in ultrafast lasers, communications, and fiber-based delivery of high-power laser beams.
 
\end{abstract}

\maketitle

\newpage

\section*{Introduction}

Starting with the visionary paper of Charles Kao in 1966 \cite{Kao1,Kao2}, the development of single-mode optical fibers has truly revolutionized optics, enabling optical telecommunication over trans-continental distances \cite{Roadmap}, fiber-based optical combs \cite{Comb} for precision metrology as well as ultrafast fiber lasers \cite{Fermann}. Giving rise to the formation of optical solitons \cite{Akhmediev}, the all-optical Kerr nonlinearity plays a pivotal role in the latter two applications of fibers. At the same time, however, excessive nonlinearities also limit data capacity in fiber transmission lines \cite{capacity} as well as pulse energies and peak powers in fiber laser systems. Mitigation of these effects has been demonstrated in large-mode area fibers \cite{Stutzki}, which nevertheless require rather sophisticated photonic crystal fiber designs to scale the mode-field diameter of the fundamental mode to 50 times the laser wavelength and beyond. Still, beam quality and non-ideal ${\rm M}^2$ factors remain a limiting problem in high-power fiber lasers \cite{Limpert}. More recently, nonlinear beam self-cleaning (BSC) has been observed as an alternative approach in simply structured multimode fibers \cite{renninger2013optical,krupa2017spatial,wright2017spatiotemporal}, that is, spatially incoherent speckle patterns observed at low-power operation slowly converge into coherent near-fundamental mode patterns at elevated powers. Owing to the large number of modes involved in a multimode system, the BSC process is often discussed in a thermodynamic picture, leading to formation of a Rayleigh-Jeans statistical distribution \cite{Picozzi,pourbeyram2022direct,wright2022physics}. In the thermodynamic approach, the fiber is treated as a closed system, and dissipative mechanisms are therefore neglected. As any useful BSC would ideally concentrate all energy in the fundamental mode, entropy inevitably has to converge to the minimum possible value of zero, that is, such a decrease of entropy defies the second law of thermodynamics. 
Here we offer new experimental evidence to clarify the physical mechanisms behind BSC. Using a previously unexplored fiber architecture, we observe BSC with unprecedented 92\% of energy contents in a single fiber mode. This feat is accompanied by a 60\% entropy drop, exceeding all previous observations by a large factor. We find that BSC may go through two different phases, with an initial convergence towards a Gaussian distribution and a rather insignificant accompanying decrease of entropy. This thermalization phase has likely been described before \cite{Picozzi,pourbeyram2022direct,wright2022physics}. However, in fibers with a dissipative core, previously unreported and much more efficient BSC appears when a second threshold is exceeded. In this second BSC phase, the beam collapses into a Townes profile with a beam diameter that falls significantly below the fundamental mode. In contrast to the early thermalization phase, this process is accompanied by a sudden decrease of the chemical potential and is therefore best understood as a phase transition or condensation, similar to Bose-Einstein condensation (BEC) \cite{BEC1,BEC2}, where the formation of spatial solitons of the Gross-Pitaevskii Equation (GPE) has been reported before \cite{GPE1,GPE2,anderson,Ketterle,Conti}. In this light, dissipative effects emerge as a crucial ingredient to meaningful BSC rather than representing a disturbing nuisance.

\section{Self-cleaning in dissipative dual-core fibers}

\begin{figure}[tbh]
\centering
\includegraphics[width=0.8 \linewidth]{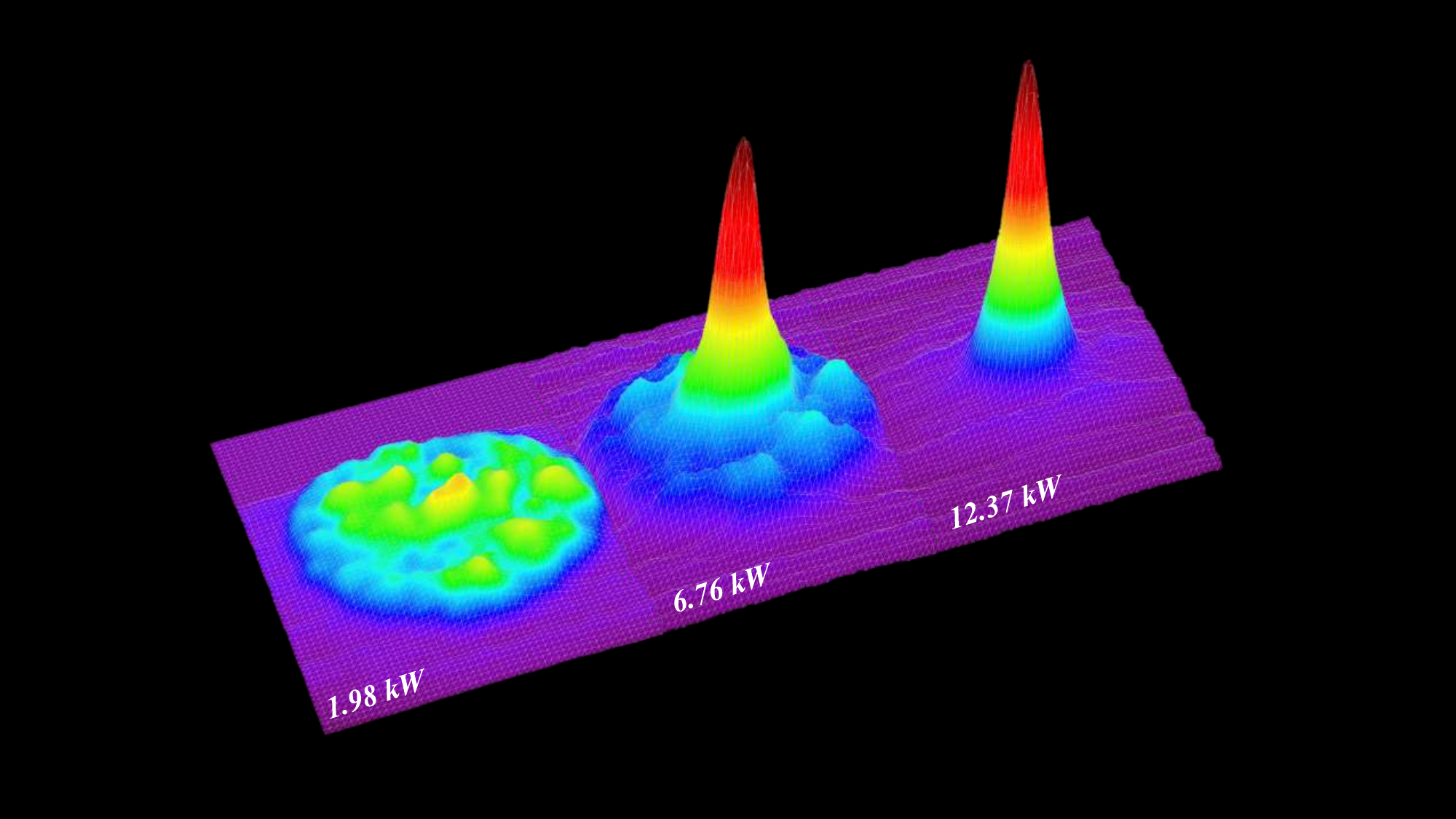}
\caption{Experimental observation of Bose-Einstein condensation during beam self-cleaning in a dissipative dual-core fiber. {\bf a}, at low powers, a characteristic incoherent speckle pattern is observed. {\bf b}, condensation sets in, concentrating the fluence in the center of the inner core. {\bf c},  final condensation into Townes profile, with nearly vanishing fluence in the pedestal.} 
\label{fig:Beamprofile}
\end{figure}

Our experiments on the BSC phenomenon employ two types of dual-core fibers with nearly identical optical properties, yet differently doped inner cores, see Methods section. These dual-core fibers are typically used in diode-pumped fiber laser amplifiers, and we are not aware of any previous experiments on BSC in this fiber type or, more generally, in fibers with dissipative cores. Figure \ref{fig:Beamprofile} shows three beam profile measurements in such a fiber for a range of input powers, clearly resembling the phase transition of transverse momentum distributions observed in dilute gases  \cite{anderson,Grimm} or plasmonic lattices \cite{plasmonic} during Bose-Einstein condensation. Three-dimensional plots reporting the phase transition have become the hallmark of the BEC community \cite{Ketterle}.

\begin{figure}[tbh]
\centering
\includegraphics[width=0.72 \linewidth]{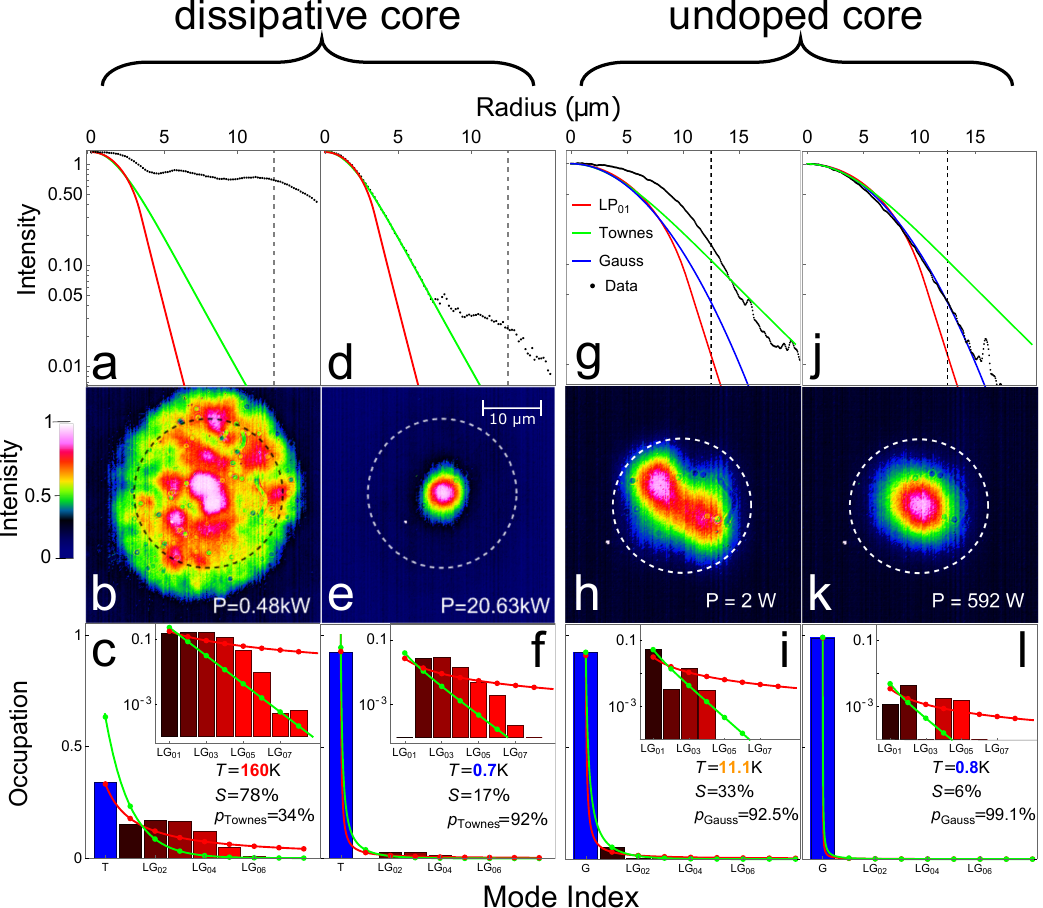}
\caption{Mode decomposition analysis for the beam self-cleaning process. {\bf a,d,g,j} azimuthally averaged near-field beam profiles extracted from beam profile measurements {\bf b,e,h,k}, respectively. Different functions have been fitted to measured radial beam profiles, indicated by red, green, and blue curves.  The core radius of 12.5 $\mu$m is indicated as a dashed circle in the beam profiles. {\bf c,f,i,l} Decomposition into Townes mode and higher-order Gauss-Laguerre residuals ${\rm LG}_{01}$ -- ${\rm LG}_{08}$ on a linear scale and a log scale (insets). Green curves indicate a fitted Bose-Einstein distribution; red curves a Rayleigh-Jeans fit. Temperatures $T$ are shown as extracted from the chemical potential $\mu/k_{\rm B}$ of the Bose-Einstein fits. $S$ indicates Gibbs entropy relative to the maximum possible value of ${\rm ln}\,9$ and $p_{\rm Townes}$ the occupation of the Townes mode. {\bf a-c} Measurement at minimum power and {\bf d-f} at maximum power with dissipative core. {\bf g-i} Measurement at highest temperature and entropy and {\bf j-l} at lowest respective values with an undoped core not showing absorption. }   
\label{fig:Decomposition}
\end{figure}

In Fig.~\ref{fig:Decomposition} we will now provide a deeper analysis of the measured beam profiles, starting with the dissipative fiber. More complete analysis of measured beam profiles is available in the SI. 2D plots of the low-power speckle pattern and the high-power Townes soliton are shown in Figs.~\ref{fig:Decomposition}{\bf b} and {\bf e}, respectively. The inner core with its 25\,$\mu$m diameter is additionally indicated by a dashed curve. Similar to previous observations of BSC \cite{deliancourt2019wavefront,krupa2017spatial}, interference of modes with different propagation constants causes speckle-like beam profiles of low spatial coherence at low input powers. Increasing the input power, nonlinear optical effects set in, and the energy contents start to confine within the inner half of the fiber core, eventually decreasing to a diameter of about 8\,$\mu$m at output powers close to the maximum power available from our laser. The latter mode field diameter already falls well below expectations from linear optics \cite{LoveSnyder}, which predict a ${\rm LP}_{01}$ mode diameter of 13\,$\mu$m, that is, $60\%$ larger than actually measured. Even scaling the ${\rm LP}_{01}$ mode to a smaller diameter, we cannot obtain a good fit to the measured beam profiles, cf. red curve in Fig.~\ref{fig:Decomposition}{\bf d}. Specifically, the intensity rolls off much slower with increasing radius than expected for an  ${\rm LP}_{01}$ mode and fits nearly perfectly to a Townes profile, cf. green curve in Fig. \ref{fig:Decomposition}{\bf d}.

We further analyzed the beam profiles using a pseudo mode expansion into orthonormal Gauss-Laguerre modes (Figs.~\ref{fig:Decomposition}{\bf c} and {\bf f}). Our analysis indicates high-power convergence against a circularly symmetric bell-shaped spatial profile, which, apart from the unexpectedly small diameter, falls off much slower at large radius than anticipated from the fundamental ${\rm LP }_{01}$ mode of the inner core. These findings call for a strong role of nonlinear optical effects, namely self-focusing. Inclusion of the latter nonlinearity in the Helmholtz Equation formally leads to the Gross-Pitaevskii-Equation (GPE), which has also been frequently used to describe soliton formation during BEC \cite{BEC1,BEC2}. The circularly symmetric eigensolutions of the GPE are known as Townes solitons, see Methods section for further details. Based on our pseudo mode expansion, we further extract the overlap of measured beam profiles with the Townes soliton and find a remarkably high value of 92\% at the highest peak power level of $\approx 21\,$kW. While a Gaussian also results in an overlap of about 80\%, the Townes profile fits nearly perfectly for intensity levels down to 5\% of the peak, that is, the remaining 8\% non-Townes contents can be nearly exclusively attributed to a pedestal region in the outer part of the core.

\begin{figure}[tbh]
\centering
\includegraphics[width=0.8 \linewidth]{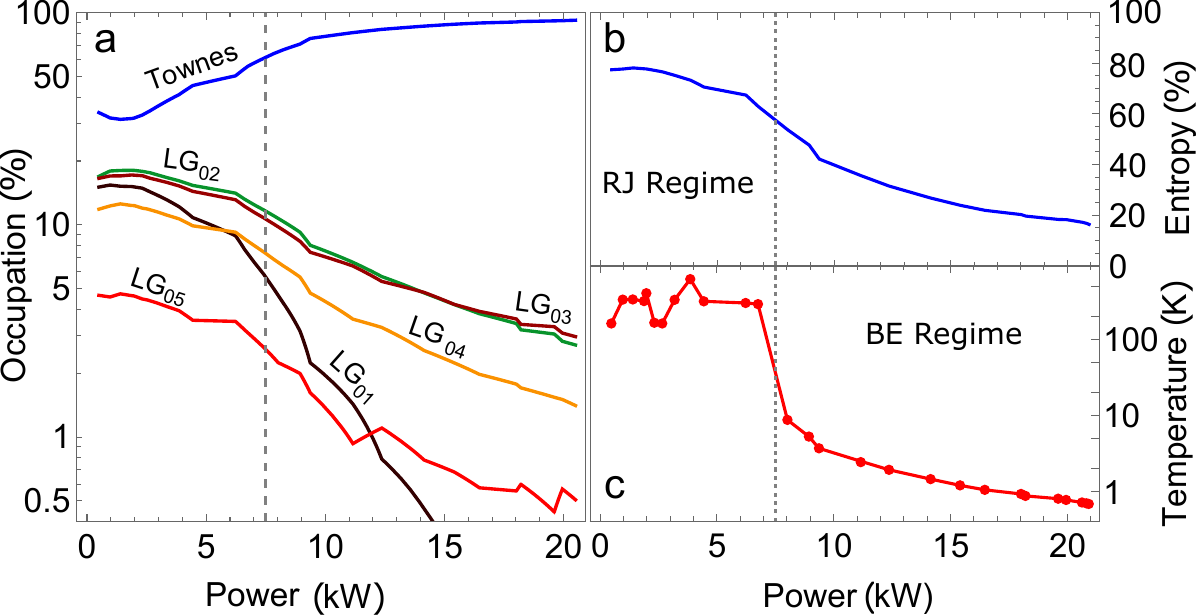}
\caption{Thermodynamic parameters extracted from the analysis of the dissipative fiber data in Fig.\ref{fig:Decomposition}. {\bf a} Modal occupation of Townes mode and 8 higher-order Gauss-Laguerre residuals ${\rm LG}_{01}$ -- ${\rm LG}_{08}$. {\bf b} Relative Gibbs entropy derived from  {\bf a}. {\bf c} Temperature $T=\mu/k_{\rm B}$ extracted from Bose-Einstein fits in Fig.\ref{fig:Decomposition}. The appearance of the phase transition is indicated by a dashed line. Initial thermalization is best described by a Rayleigh-Jeans distribution; cooling after condensation by Bose-Einstein statistics.}   
\label{fig:thermo results}
\end{figure}

Our analysis now gives us access to a full thermodynamic analysis, see Fig.~\ref{fig:Decomposition}{\bf c,f,i,l} and Fig.~\ref{fig:thermo results}. First, computing Gibbs entropy $S$ from the modal decomposition, we observe a clear decrease from 77\% to 17\% of the maximum possible entropy value $S_{\rm max}={\rm ln}\,9$ in our system, that is, when energy is equally distributed over all 9 modes in our expansion. Using a Boltzmann entropy definition instead, the decrease is less dramatic, yet still seems to be significant (cf.~discussion in the SI). For this type of fiber, an entropy decrease does not appear surprising as our fiber core is dissipative. Therefore, the doped fiber must not be treated as a closed thermodynamic system, other than what was assumed in previous theoretical explanations of the BSC phenomenon \cite{pourbeyram2022direct}. Moreover, one can now fit different thermal distributions to the modal decompositions, which indicate the appearance of a phase transition at about 7.5\,kW peak power. While modal decompositions and entropy do not show any discontinuous behavior at this threshold power (Fig.~\ref{fig:thermo results}{\bf a,b}), the temperature $T=\mu/k_{\rm B}$ exhibits a sharp transition from low-power temperatures of several hundred Kelvin to a few Kelvin during the condensation process (Fig.~\ref{fig:thermo results} {\bf c}), similar to the sudden decrease of the chemical potential $\mu$ at the condensation point in BEC. We only observe this phase transition when fitting Bose-Einstein statistics \cite{Zitelli} to our decompositions while Rayleigh-Jeans distributions tend to provide a better fit to the data below the transition point. We therefore distinguish an initial thermalization phase with increasing accumulation of energy in the fundamental mode at roughly constant temperatures and a cooling phase that starts with the condensation process above $P \approx 7.5\,$kW. In the former phase, one only observes an insignificant decrease of entropy, which nevertheless heavily accelerates beyond the phase transition.

\section{Comparison to conservative fibers}

\begin{figure}[tbh]
\centering
\includegraphics[width=0.8 \linewidth]{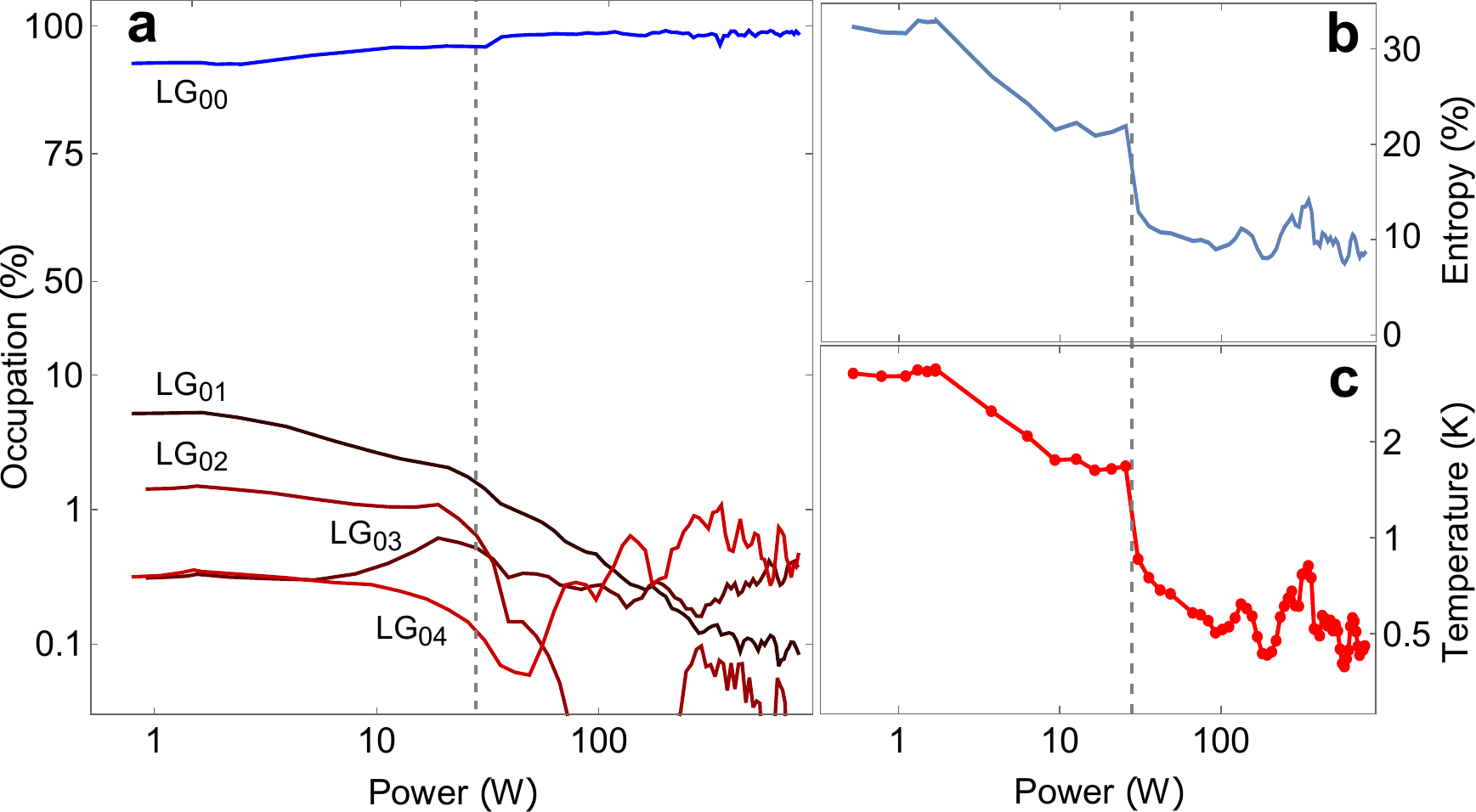}
\caption{Thermodynamic parameters extracted from the analysis of the undoped fiber data in Fig.\ref{fig:Decomposition}. {\bf a} Modal occupation Gauss-Laguerre modes ${\rm LG}_{00}$ -- ${\rm LG}_{04}$. {\bf b} Relative Gibbs entropy derived from  {\bf a}. {\bf c} Temperature $T=\mu/k_{\rm B}$ extracted from Bose-Einstein fits in Fig.\ref{fig:Decomposition}. The appearance of potential phase transition is indicated by a dashed line.}     
\label{fig:undoped}
\end{figure}

In order to better understand the role of dissipative effects in BSC, we performed a series of reference measurements in an undoped fiber with otherwise identical geometry, see Fig.~\ref{fig:undoped}. Using this formally conservative fiber, we see a similar, yet much less pronounced BSC effect already at rather moderate peak power levels, that is, at 250 times lower intensities compared to the experiment with the doped fiber, see Fig.\ref{fig:Decomposition}{\bf h,k}, where we display the beam profiles that gave rise to extreme temperatures and entropies in our subsequent analysis. Further increase of powers neither gives rise to beam profile break-ups nor to any improvement of BSC. In order to see any BSC effect in this step-index fiber, we changed focusing parameters at the input for obtaining profiles with $>90\%$ Gaussian contents already at extremely low powers. Increasing the power to about 0.5\,kW, the Gaussian contents would then increase to $>99\%$. Specifically, no convergence toward the fundamental ${\rm LP}_{01}$ mode of the step-index fiber is observed, and the azimuthally averaged low-power beam profile is well represented by a Gaussian distribution as in classical Maxwell-Boltzmann theory, cf.~SI. We take the latter observation as an indication that nonlinearities probably play a minor role in this power range and that radial beam profiles can be represented by linear Gaussian statistics. Figure \ref{fig:undoped} shows further analysis of the measurements conducted with the undoped fiber. Specifically, Gibbs entropy decreases also in these measurements (Fig. \ref{fig:undoped}{\bf b}), which may appear to contradict the assumption of a closed system. Using a Boltzmann definition instead, however, entropy appears to be rather constant, see SI for further details. We therefore suspect that there also may be a hidden dissipative mechanism in the undoped fiber. Moreover, we also observe a small discontinuity in the chemical potential, see Fig.  \ref{fig:undoped} {\bf c}, which might be interpreted as an indication for the onset of a condensation process. Further research appears necessary to clarify the behavior of optical thermodynamic systems in the thermalization phase.

\begin{figure}[t]
\centering
\includegraphics[width=0.7 \linewidth]{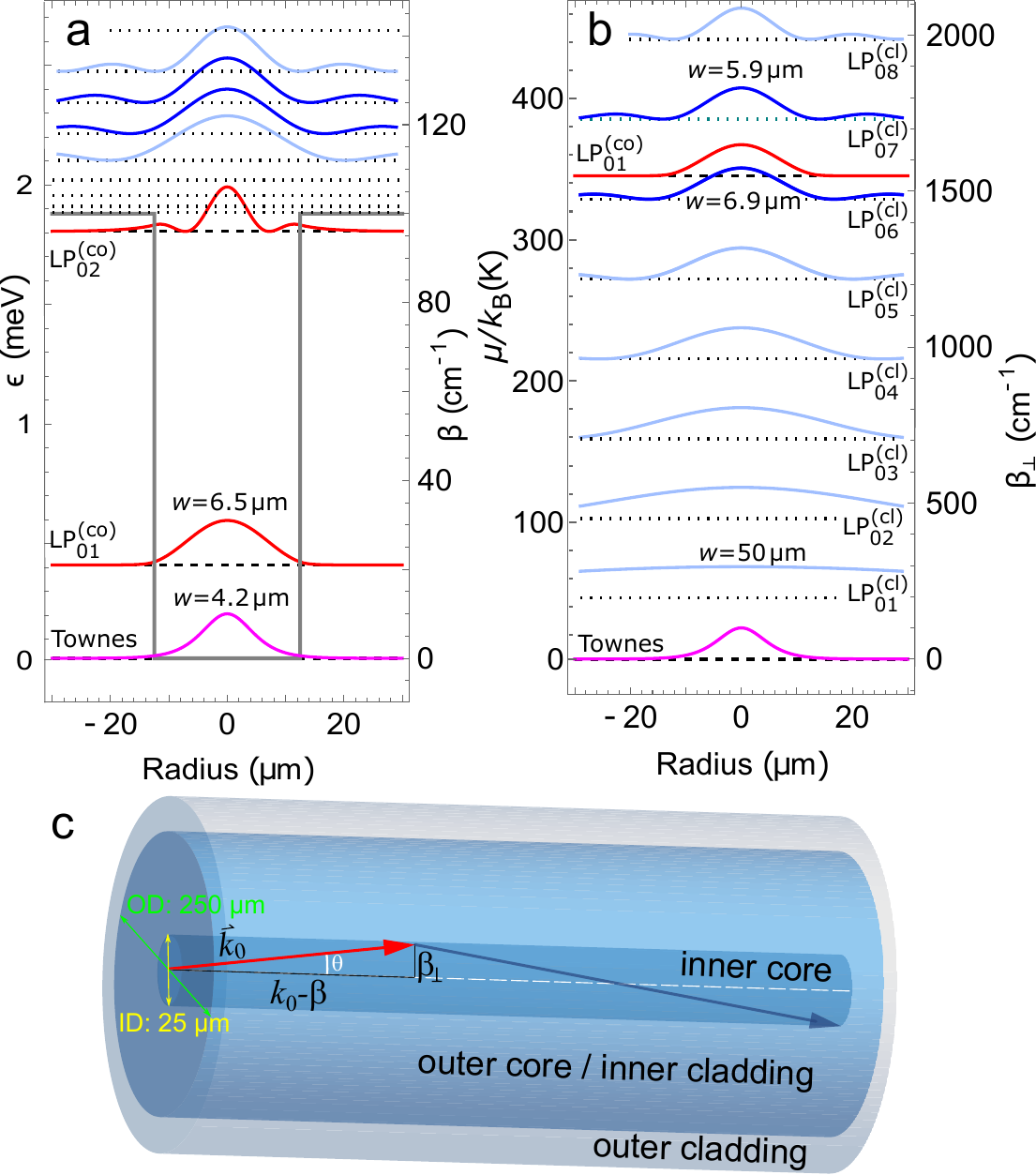}
\caption{Computed eigenmodes \cite{LoveSnyder} of the inner and outer core, ${\rm LP}_{0x}^{\rm (co)}$ and ${\rm LP}_{0x}^{\rm (cl)}$, respectively.  {\bf a} Vertical offset adjusted relative to propagation constants $\beta$ of the of the most relevant four-wave mixing interacting modes. Left vertical axis indicates the equivalent energy $\epsilon$ in a trap potential (gray lines).  {\bf b} Same with vertical offsets adjusted according to perpendicular wavenumbers $\beta_\perp$. Left axis shows the equivalent chemical potential $\mu$, which is expressed in Kelvin units. The magenta curve additionally indicates the Townes profile, which arises at absolute zero temperature and energy. {\bf c} Schematic picture of the dual core fiber, indicating the Pythagorean relationship between wavevector $k_0$, perpendicular wave vector component $\beta_\perp$, and propagation constant $\beta$, which is the eigenvalue solution of the transverse Helmholtz equation.}
\label{fig:Levels}
\end{figure}

\section{Discussion}
Beam self-cleaning obviously relies on four-wave mixing (FWM) processes in the multimode fiber. Typically, one only discusses FWM between different wavelength components of a broadband spectrum in nonlinear fiber optics. In a multimode environement, however, FWM can additionally transfer energy from one mode to another even when all wavelengths involved are identical. To this end, it is helpful to understand the modal structure of our dual-core system, which is shown in Fig. \ref{fig:Levels}{\bf a} ordered by the respective propagation constant $\beta$. Using the relationship $\epsilon=h c \beta$, one can formally convert propagation constants into energies $\epsilon$, with the speed of light $c$  and Planck's constant $h$. Rescaling to energy units, we can identify two modes, namely the ${\rm LP}_{01}$ and ${\rm LP}_{02}$ mode within the 1.9\,meV trap depth of the inner core. The latter mode is already extremely close to the modal cutoff. As the fiber was wound on a spool in the experiments, the ${\rm LP}_{02}$ mode is therefore expected to experience strong guiding losses into the quasi-continuum of modes in the outer core at $\epsilon>1.9\,$meV, which we will refer to as cladding modes in the following. With increasing self-focusing nonlinearity, the fundamental ${\rm LP}_{01}$ mode starts to shrink and converges towards the Townes profile. As diffractive and self-focusing effects ideally compensate each other in the convergence limit, the Townes mode propagates at the phase velocity of the core material, i.e., $\beta_{\rm Townes}=\epsilon_{\rm Townes}=0$. 

\begin{figure}[t]
\centering
\includegraphics[width=0.7 \linewidth]{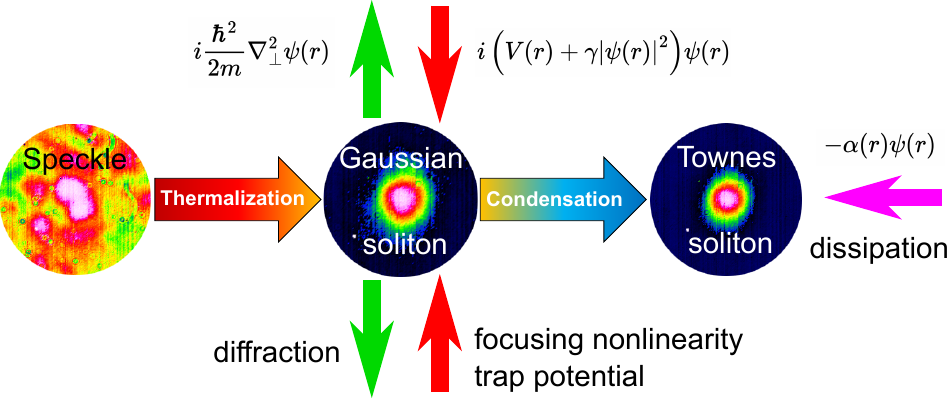}
\caption{Schematic representation of the various processes in the beam self-cleaning process in step-index fibers. In linear optics, diffraction is compensated by the trap potentials, which leads to the formation of the fundamental ${\rm LP}_{00}$ eigenmode. According to our experimental observations, additional self-focusing effects lead to formation of a Gaussian eigenmode, which is similar to previous explanations of the self-cleaning effect. Much more effective cleaning can be observed in the presence of additional dissipative effects. In the latter case, the beam profile condenses in a Townes state, with a temperature equivalent of 0\,K.}
\label{fig:theo}
\end{figure}

For an understanding of nonlinear interaction between these modes, it is further helpful to employ the ray analysis approach outlined in \cite{Crenn,LoveSnyder}, exploiting a Pythagorean relationship $k_0^2=(k_0-\beta)^2+\beta_\perp^2$ between wavenumber $k_0$ in the bulk core material, propagation constant $\beta$, and a transverse wavenumber component $\beta_\perp\approx \sqrt{2 k_0 \beta}$. The geometric construction behind this relationship is depicted in Fig.~\ref{fig:Levels}{\bf c}. Reordering the modes by $\beta_\perp$ (Fig.~\ref{fig:Levels}{\bf b}), one immediately understands that core and cladding modes with similar $\beta_\perp$ display strong overlap and therefore interact most efficiently in the four-wave mixing process. Already at low powers, therefore, there appears a relatively strong interaction between cladding modes ${\rm LP}_{06,07}$ and the fundamental core mode ${\rm LP}_{01}$. Consequently, initial population in the ${\rm LP}_{06,07}$ modes can effectively be trapped in the core potential, giving rise to an initial thermalization process. Once significant energy has been trapped inside the core, FWM may also convert photons in the fundamental core mode back into cladding modes outside the resonant band around  ${\rm LP}_{06,07}$. This back conversion would usually limit further accumulation of energy in the trap state and may also explain why we do not see convergence towards the fundamental  ${\rm LP}_{01}$ core mode but rather a statistical distribution for the undoped fiber. Moreover, while FWM processes themselves are always conservative, they can convey energy to outside our observation window, i.e., from the inner to the outer core. Therefore FWM may effectively cause a nonlinear loss mechanism in our fiber. Consequently, our multimode system cannot be considered a closed thermodynamic system, and entropy may very well decrease without violating the second law of thermodynamics, i.e., even in a virtually conservative fiber.

In the wavelength degenerate FWM picture, the transverse wavenumber $\beta_\perp$ emerges as the conserved quantity rather than $\epsilon$. Converting $\beta_\perp$ into energy units and further to a temperature, we can find a connection to the chemical potential $\mu=k_B T = h c \beta_\perp$, cf. left scale of Fig.~\ref{fig:Levels}{\bf b}. The  ${\rm LP}_{01}$ core mode and neighboring ${\rm LP}_{06,07}$ cladding modes are located in a temperature range from $300$ to $400\,$K, which favorably agrees with temperatures during the thermalization phase in Fig.~\ref{fig:thermo results}{\bf b}. Once the condensation process sets in, the temperature suddenly drops to below $10\,$K and slowly further decreases into the milli-Kelvin regime. In contrast, the lowest-lying ${\rm LP}_{01}$ cladding mode is located at a temperature equivalent of $45\,$K, which is nearly 2 orders of magnitude above the final temperature of the condensate. Such low temperatures can only be explained by population of the Townes mode, i.e., the solution of the Gross-Pitaevskii-Equation for eigenvalue 0.

\section{Conclusions}

Our analysis has shown that the BSC phenomenon can be understood using two complementary pictures. In the thermodynamic approach, the beam profile undergoes an initial thermalization process, leading to Gaussian beam profiles, similar to what would be expected in classical Maxwell-Boltzmann statistics. In the energy-related picture of Fig.~\ref{fig:Levels}, the thermalization process can be understood as accumulation of photons in the trap potential. In this phase, temperatures cannot go much below $\approx 300\,$K, which is confirmed by the extracted chemical potentials in  Fig.~\ref{fig:thermo results}{\bf b}. FWM-mediated losses during the thermalization phase can be understood in a similar fashion, e.g., as evaporative cooling in BEC, which selectively removes particles with highest energy from the distribution. While evaporative cooling is the key mechanism for obtaining BEC in dilute gases, it is apparently insufficient to induce condensation in our optical thermodynamic system, which requires an additional dissipative mechanism to allow condensation into the Townes state. This final transition toward near-zero temperatures is characterized by a sudden drop of chemical potential, which is also characteristic for BEC. This finding confirms the presence of a phase transition in our experiments with the dissipative fiber. This phase transition contrasts the gradual cooling or thermalization processes reported in previous literature, e.g., \cite{pourbeyram2022direct,mangini}. 

Alternatively, one can now also interpret the condensation process in the language of nonlinear optics. After a sufficient number of photons has been accumulated in the trap state, FWM processes kick in, populating cladding states ${\rm LP}_{07}$ and higher. Consequently, the beam starts to contract due to self-focusing. This process is inevitably accompanied by a transfer of photons into cladding modes that cannot be guided in the inner core anymore. This nonlinear loss mechanism can be included in the GPE as an imaginary part of the parameter $\gamma$, i.e., similar to the action of the effective saturable absorber in Kerr-lens mode-locking \cite{Haus2}.
Without suitable stabilization, the beam profile is expected to undergo a filamentary breakup \cite{Bespalov}, but apparently such breakup is prevented by the combination of the trap potential and dissipative mechanisms, cf.~Fig.~\ref{fig:theo}, explaining the competing action of reactive and dissipative beam shaping mechanisms. If the latter mechanisms are restricted to FWM-mediated transfer of energy into the continuum, we observe the formation of a Gaussian soliton, i.e., the eigenvalue solution of the Helmholtz equation in free space. Intensity is then automatically limited by the nonlinear loss mechanism, which probably arrests the beam collapse. If linear dissipative mechanisms dominate, however, the self-focusing action has to be taken into the equation [Fig.~\ref{fig:Levels}{\bf c}]. In this case, the Helmholtz equation becomes the Gross-Pitaevskii Equation (GPE), famously known for explaining soliton formation in Bose-Einstein condensates \cite{anderson,Grimm}. The eigenvalue solution of the GPE is the Townes soliton, which is also known to be unstable due to modulation instability \cite{Bespalov}, requiring a dissipative effect for regularization \cite{Malomed}. Strictly speaking, therefore, the Townes soliton of the GPE is actually a dissipative soliton \cite{Akhmediev}, and dissipative effects appear to play a similar stabilizing role as saturable absorption does for the mode-locking process \cite{Haus2}.

The previously underestimated stabilizing role of dissipation therefore sheds new light on the BSC phenomenon and opens an avenue for a plethora of novel applications in high-power fiber lasers, flexible optical beam delivery, and optical pulse compression in multimode fibers. In our work, we exploited dissipative BSC yielding near-perfect $>90\%$ self-cleaning, which exceeds all previous demonstrations by a large factor. Admittedly, this feat currently comes with rather large overall losses, which can nevertheless be strongly reduced when the dissipative effect is limited to an annular zone to sufficiently delay the onset of modulation instability, similar to what was suggested in Bose-Einstein condensation\cite{Malomed}. Such specifically designed cleaning fibers may resolve one of the plaguing problems of high-power fiber lasers, namely their often limited beam quality. Dissipative beam cleaning also enables applications that are currently still impossible because of Kerr-induced bottlenecks. Such cleaning fibers may therefore further revolutionize the field of fiber lasers, similar to the impact of evaporative cooling in Bose-Einstein condensation.

\section*{Methods}
\subsection*{Laser source}
The experiments employed a home-built femtosecond fiber laser with 1560\,nm central wavelength,  about 20\,nm bandwidth (FWHM), delivering pulses with up to 21\,kW peak power, which can be compressed down to 300\,fs pulse duration, yielding a near ideal time-bandwidth product of 0.34. A combination of a half-wave retarder and a Faraday isolator was used for variable attenuation of the laser source.
\subsection*{Fiber parameters}
The BSC experiments were conducted in a 3\,m long dual-core fiber (iXblue IXF-2CF-EY-PM-25-250-AF) with 25 and 250\,$\mu$m diameter of the inner core and the cladding (or outer core), respectively. The polarization maintaining inner core of this fiber is co-doped with Ytterbium and Erbium, giving rise to dissipative losses of about 20 dB/m at 1560 nm. The birefringence oft he fiber was specified as $\ge 0.5 \times 10^{-4}$. For comparison, a purely passive fiber (iXblue IXF-2CF-PAS-PM-25-250-0.08) with otherwise identical optical properties was used, but did not show pronounced BSC. Given the rather small NA=0.085 of the inner core of these fibers, the LP$_{01}^{\rm (co)}$ is very close to the cutoff. As the fibers were wound on a spool with 30 cm diameter, the central core essentially only guides the LP$_{01}^{\rm (co)}$ mode ($w=6.5\,\mu$m, $\beta=20.6\,$cm$^{-1}$). The outer core (NA$=0.45$) nominally guides a total of 73 LP$_{0n}^{\rm (cl)}$ modes. 

%Among this quasi-continuum, modes LP$_{05}^{\rm (co)}$ and LP$_{06}^{\rm (co)}$ exhibit similar %effective areas and $\beta_\perp$ as the fundamental core mode LP$_{01}^{\rm (co)}$, which is %expected to lead to highly efficient four-wave mixing interactions between these three modes.
\subsection*{Data analysis and pseudomode decomposition}
The frames measured at highest powers were first fitted to an elliptical Gaussian 
\begin{equation}
\propto \exp\left[ -(x-x_0)^2/w_x^2 -(y-y_0)^2/w_y^2 \right]
\end{equation}
to determine the center position $(x_0,y_0)$ and the ellipticity
parameter $\epsilon=|1 - w_x/w_y|\approx 13.5\%$ of the birefringent core. Data is then resampled using elliptical coordinates $(r\cos(\varphi)/w_x,r\sin(\varphi)/w_y)$, yielding azimuthally averaged reduced data sets $\overline{I(r,P)}$. As the extracted $\overline{I(r,P)}$ do not exhibit any apparent field nodes within the core radius, we analyze the radial beam profiles with a pseudo-mode expansion. As the underlying truncated Bessel $J_0$ functions of the LP modes are not orthonormal, we choose the Gauss-Laguerre modes LG$_{0n}$ as a more convenient basis for our analysis, yielding mode amplitudes $a_n$, $n=0 \dots 8$. We further employ a modal expansion of the Townes soliton to determine the solitonic contents from our beam profiles. Normalized sequences $|a_n|^2$ are then fitted to Maxwell, Rayleigh-Jeans, and Bose-Einstein distributions, yielding temperature and chemical potential estimates as well as a goodness of fit for judging suitability of the respective model. Moreover, interpreting $p_n=|a_n|^2$ allows to determine the entropy via $S=-\sum_n p_n \log p_n$.
\subsection*{The Gross-Pitaevskii Equation (GPE)}
The GPE describes the ground state $\psi(r)$ of an interacting system of identical bosons with the Hamiltonian
\begin{equation}
\hat{H}= i \left(\frac{\hbar^2}{2 m} \nabla_\perp^2 + V(r) + \gamma \left| \psi(r) \right|^2 \right) \label{eq:GPE1}
\end{equation}
%\begin{eqnarray}
%i \frac{\hbar^2}{2 m} \nabla_\perp^2 \psi(r) \\ i \left( V(r) + \gamma \left| \psi(r) \right|^2 \right) \psi(r) \\ - \alpha(r) \psi(r)
%\end{eqnarray}
in a binding potential $V(r)$. Here a positive value of $\gamma$ accounts for an attractive nonlinearity of the system whereas negative values model repulsive collision processes. In a gradient index fiber, the kinetic energy operator $\nabla_\perp^2 \hbar^2/2 m$ is interchanged for $\nabla_\perp^2/2 \beta_0$, nonlinearity is caused by the Kerr effect $\gamma= \beta_0 n_2$, and $V(r)=\beta_0 n(r)$ describes the step index profile 
\begin{equation}
n(r)= \left\{
\begin{array}{ll}
      n_{\rm co} & r\leq a_0 \\
      n_{\rm cl} & r > a_0 \\
\end{array}
\right. 
\end{equation}
of the fiber. Here $n_{\rm co}$ and $n_{\rm cl}$ are the inner and outer core refractive indices, respectively, and $a_0$ is the radius of the inner core. The GPE then takes the form of the nonlinear paraxial wave equation
\begin{equation}
2 i \beta_0 \frac{\partial A(r,z)}{\partial z} = - \left[\nabla_\perp^2 + \beta_0^2 \left(V(r) + n_2 \left| A(r,z) \right|^2 \right) \right] A(r,z).   \label{eq:GPE2}
\end{equation}
\subsection*{The Soliton of the GPE}
Using a normalized version of the Hamiltonian of the GPE with unit step potential $V(r)$
\begin{equation}
\hat{G} = \frac{1}{r} \partial_r r \partial_r + \delta V(r) + \gamma |u|^2 \label{eq:GPE3}
\end{equation}
the eigenvalue problem $\hat{G} u(r) = 0$ can be satisfied for $\delta=-1.13$ and $\gamma=5.5$, i.e., diffraction, nonlinearity, and the parabolic potential exactly balance each other, giving rise to a Townes soliton $u(r)$ propagating exactly at the speed of light in the core material of the fiber, that is, the propagation constant $\beta$ is exactly zero. Adding a soliton phase $\psi_0$ to the Hamiltonian $\hat{G}$, a continuous range of soliton solutions can be found for the range $\gamma=[0,5.5]$.
%\subsection*{Numerical Simulations}
%Numerical simulations have been conducted based on Eq.~(\ref{eq:GPE3}) with $\gamma$ and %$\delta$ supporting formation of a Townes soliton. The Laplacian has been computed with an 7th %order Hermitian interpolation of $u(r)$, and Eq.~(\ref{eq:GPE3}) was integrated out using a %backward differentiation formula (BDF) within the Mathematica NDSolve environment.
 
%In order to ensure numerical stability of the code, a dissipative coating has been assumed at $r>a_0$, i.e., any light entering the cladding is considered lost
%\begin{equation}
%V(r)=  \left\{
%\begin{array}{ll}
%      \Delta r^2 & r\leq a_0 \\
%      \Delta a_0^2 + i \alpha & r > a_0 \\
%\end{array} \right.
%\end{equation}
%Choosing a regularization parameter $\alpha=10^4$ stabilizes the code for most situations, however, needle-like artifacts may still appear in certain situations. Such artifacts indicate a breakdown of the paraxial approximation underlying Eqs.~(\ref{eq:GPE1}-\ref{eq:GPE3}). In contrast, the Townes soliton propagates at constant shape with less than $10^{-3}$ deviations from energy conservation.
\section*{Funding}
National Key R\&D Program of China (Grant 2023YFB3611000), National Natural Science Foundation of China (NSFC) (Grants: 62105237, 62227821, 62205015 and 62275015)

\section*{Acknowledgments}
GS acknowledges fruitful discussions with Stefan Wabnitz and Mario Zitelli (Sapienza University of Rome), Demetrios Christodoulides (University of Southern California), Fan Wu (CREOL), Frank Wise (Cornell University), and Pavel Sidorenko (Technion).

\section*{Author contributions}
J.F. and M.H. conceived the experiment. J.Z. collected all experimental data. G.S. performed the modal decompositions and further data analysis. C.M. conducted numerical simulations. J.F. and G.S. wrote the manuscript with feedback from all authors.

\section*{Competing interest}
The authors declare no competing interests.

\section*{Data availability}
All data and codes to generate the figures are available upon request. 

\newpage
\bibliographystyle{naturemag}
\bibliography{Ref}

\begin{thebibliography}{10}
\expandafter\ifx\csname url\endcsname\relax
  \def\url#1{\texttt{#1}}\fi
\expandafter\ifx\csname urlprefix\endcsname\relax\def\urlprefix{URL }\fi
\providecommand{\bibinfo}[2]{#2}
\providecommand{\eprint}[2][]{\url{#2}}

\bibitem{Kao1}
\bibinfo{author}{Kao, K.~C.} \& \bibinfo{author}{Hockham, G.~A.}
\newblock \bibinfo{title}{Dielectric-fibre surface waveguides for optical frequencies}.
\newblock \emph{\bibinfo{journal}{Proc. IEE}} \textbf{\bibinfo{volume}{113}}, \bibinfo{pages}{1151--1158} (\bibinfo{year}{1966}).

\bibitem{Kao2}
\bibinfo{author}{Kao, C.~K.}
\newblock \bibinfo{title}{Nobel lecture: Sand from centuries past: Send future voices fast}.
\newblock \emph{\bibinfo{journal}{Rev. Mod. Phys.}} \textbf{\bibinfo{volume}{82}}, \bibinfo{pages}{2299--2303} (\bibinfo{year}{2010}).

\bibitem{Roadmap}
\bibinfo{author}{Agrell, E.} \emph{et~al.}
\newblock \bibinfo{title}{Roadmap of optical communications}.
\newblock \emph{\bibinfo{journal}{J. Opt.}} \textbf{\bibinfo{volume}{18}} (\bibinfo{year}{2016}).

\bibitem{Comb}
\bibinfo{author}{Holzwarth, R.} \emph{et~al.}
\newblock \bibinfo{title}{Optical frequency synthesizer for precision spectroscopy}.
\newblock \emph{\bibinfo{journal}{Phys. Rev. Lett}} \textbf{\bibinfo{volume}{85}}, \bibinfo{pages}{2264--2267} (\bibinfo{year}{2000}).

\bibitem{Fermann}
\bibinfo{author}{Fermann, M.~E.} \& \bibinfo{author}{Hartl, I.}
\newblock \bibinfo{title}{Ultrafast fibre lasers}.
\newblock \emph{\bibinfo{journal}{Nature Photon.}} \textbf{\bibinfo{volume}{7}}, \bibinfo{pages}{868--874} (\bibinfo{year}{2013}).

\bibitem{Akhmediev}
\bibinfo{author}{Grelu, P.} \& \bibinfo{author}{Akhmediev, N.}
\newblock \bibinfo{title}{Dissipative solitons for mode-locked lasers}.
\newblock \emph{\bibinfo{journal}{Nature Photon.}} \textbf{\bibinfo{volume}{6}}, \bibinfo{pages}{84--92} (\bibinfo{year}{2012}).

\bibitem{capacity}
\bibinfo{author}{Temprana, E.} \emph{et~al.}
\newblock \bibinfo{title}{Overcoming {Kerr}-induced capacity limit in optical fiber transmission}.
\newblock \emph{\bibinfo{journal}{Science}} \textbf{\bibinfo{volume}{348}}, \bibinfo{pages}{1445--1448} (\bibinfo{year}{2015}).

\bibitem{Stutzki}
\bibinfo{author}{Stutzki, F.} \emph{et~al.}
\newblock \bibinfo{title}{Designing advanced very-large-mode-area fibers for power scaling of fiber-laser systems}.
\newblock \emph{\bibinfo{journal}{Optica}} \textbf{\bibinfo{volume}{1}}, \bibinfo{pages}{233--242} (\bibinfo{year}{2014}).

\bibitem{Limpert}
\bibinfo{author}{Limpert, J.} \emph{et~al.}
\newblock \bibinfo{title}{{Yb}-doped large-pitch fibres: effective single-mode operation based on higher-order mode delocalisation}.
\newblock \emph{\bibinfo{journal}{Light Sci. Appl.}} \textbf{\bibinfo{volume}{1}}, \bibinfo{pages}{e8--e8} (\bibinfo{year}{2012}).

\bibitem{renninger2013optical}
\bibinfo{author}{Renninger, W.~H.} \& \bibinfo{author}{Wise, F.~W.}
\newblock \bibinfo{title}{Optical solitons in graded-index multimode fibres}.
\newblock \emph{\bibinfo{journal}{Nature communications}} \textbf{\bibinfo{volume}{4}}, \bibinfo{pages}{1719} (\bibinfo{year}{2013}).

\bibitem{krupa2017spatial}
\bibinfo{author}{Krupa, K.} \emph{et~al.}
\newblock \bibinfo{title}{Spatial beam self-cleaning in multimode fibres}.
\newblock \emph{\bibinfo{journal}{Nature Photonics}} \textbf{\bibinfo{volume}{11}}, \bibinfo{pages}{237--241} (\bibinfo{year}{2017}).

\bibitem{wright2017spatiotemporal}
\bibinfo{author}{Wright, L.~G.}, \bibinfo{author}{Christodoulides, D.~N.} \& \bibinfo{author}{Wise, F.~W.}
\newblock \bibinfo{title}{Spatiotemporal mode-locking in multimode fiber lasers}.
\newblock \emph{\bibinfo{journal}{Science}} \textbf{\bibinfo{volume}{358}}, \bibinfo{pages}{94--97} (\bibinfo{year}{2017}).

\bibitem{Picozzi}
\bibinfo{author}{Sun, C.} \emph{et~al.}
\newblock \bibinfo{title}{Observation of the kinetic condensation of classical waves}.
\newblock \emph{\bibinfo{journal}{Nature Phys.}} \textbf{\bibinfo{volume}{8}}, \bibinfo{pages}{470--474} (\bibinfo{year}{2012}).

\bibitem{pourbeyram2022direct}
\bibinfo{author}{Pourbeyram, H.} \emph{et~al.}
\newblock \bibinfo{title}{Direct observations of thermalization to a {Rayleigh}--{Jeans} distribution in multimode optical fibres}.
\newblock \emph{\bibinfo{journal}{Nature Physics}} \textbf{\bibinfo{volume}{18}}, \bibinfo{pages}{685--690} (\bibinfo{year}{2022}).

\bibitem{wright2022physics}
\bibinfo{author}{Wright, L.~G.}, \bibinfo{author}{Wu, F.~O.}, \bibinfo{author}{Christodoulides, D.~N.} \& \bibinfo{author}{Wise, F.~W.}
\newblock \bibinfo{title}{Physics of highly multimode nonlinear optical systems}.
\newblock \emph{\bibinfo{journal}{Nature Physics}} \textbf{\bibinfo{volume}{18}}, \bibinfo{pages}{1018--1030} (\bibinfo{year}{2022}).

\bibitem{BEC1}
\bibinfo{author}{Denschlag, J.} \emph{et~al.}
\newblock \bibinfo{title}{Generating solitons by phase engineering of a {Bose}-{Einstein} condensate}.
\newblock \emph{\bibinfo{journal}{Science}} \textbf{\bibinfo{volume}{287}}, \bibinfo{pages}{97--101} (\bibinfo{year}{2000}).

\bibitem{BEC2}
\bibinfo{author}{Strecker, K.}, \bibinfo{author}{Partridge, G.}, \bibinfo{author}{Truscott, A.} \& \bibinfo{author}{Hulet, R.}
\newblock \bibinfo{title}{Formation and propagation of matter-wave soliton trains}.
\newblock \emph{\bibinfo{journal}{Nature}} \textbf{\bibinfo{volume}{417}}, \bibinfo{pages}{150--153} (\bibinfo{year}{2002}).

\bibitem{GPE1}
\bibinfo{author}{Liang, Z.}, \bibinfo{author}{Zhang, Z.} \& \bibinfo{author}{Liu, W.}
\newblock \bibinfo{title}{Dynamics of a bright soliton in bose-einstein condensates with time-dependent atomic scattering length in an expulsive parabolic potential}.
\newblock \emph{\bibinfo{journal}{Phys. Rev. Lett.}} \textbf{\bibinfo{volume}{94}} (\bibinfo{year}{2005}).

\bibitem{GPE2}
\bibinfo{author}{Abdullaev, F.}, \bibinfo{author}{Caputo, J.}, \bibinfo{author}{Kraenkel, R.} \& \bibinfo{author}{Malomed, B.}
\newblock \bibinfo{title}{Controlling collapse in {Bose}-{Einstein} condensates by temporal modulation of the scattering length}.
\newblock \emph{\bibinfo{journal}{Phys. Rev. A}} \textbf{\bibinfo{volume}{67}} (\bibinfo{year}{2003}).

\bibitem{anderson}
\bibinfo{author}{Anderson, M.}, \bibinfo{author}{Ensher, J.}, \bibinfo{author}{Matthews, M.}, \bibinfo{author}{Wieman, C.} \& \bibinfo{author}{Cornell, E.}
\newblock \bibinfo{title}{Observation of {Bose}-{Einstein} condensation in a dilute atomic vapor}.
\newblock \emph{\bibinfo{journal}{Science}} \textbf{\bibinfo{volume}{269}}, \bibinfo{pages}{198--201} (\bibinfo{year}{1995}).

\bibitem{Ketterle}
\bibinfo{author}{Ketterle, W.}
\newblock \bibinfo{title}{Nobel lecture: When atoms behave as waves: {Bose-Einstein} condensation and the atom laser}.
\newblock \emph{\bibinfo{journal}{Rev. Mod. Phys.}} \textbf{\bibinfo{volume}{74}}, \bibinfo{pages}{1131--1151} (\bibinfo{year}{2002}).

\bibitem{Conti}
\bibinfo{author}{Conti, C.}, \bibinfo{author}{Leonetti, M.}, \bibinfo{author}{Fratalocchi, A.}, \bibinfo{author}{Angelani, L.} \& \bibinfo{author}{Ruocco, G.}
\newblock \bibinfo{title}{Condensation in disordered lasers: Theory, $3\mathrm{D}+1$ simulations, and experiments}.
\newblock \emph{\bibinfo{journal}{Phys. Rev. Lett.}} \textbf{\bibinfo{volume}{101}}, \bibinfo{pages}{143901} (\bibinfo{year}{2008}).

\bibitem{Grimm}
\bibinfo{author}{Stellmer, S.}, \bibinfo{author}{Tey, M.~K.}, \bibinfo{author}{Huang, B.}, \bibinfo{author}{Grimm, R.} \& \bibinfo{author}{Schreck, F.}
\newblock \bibinfo{title}{{Bose}-{Einstein} condensation of strontium}.
\newblock \emph{\bibinfo{journal}{Phys. Rev. Lett.}} \textbf{\bibinfo{volume}{103}} (\bibinfo{year}{2009}).

\bibitem{plasmonic}
\bibinfo{author}{Hakala, T.~K.} \emph{et~al.}
\newblock \bibinfo{title}{Bose-einstein condensation in a plasmonic lattice}.
\newblock \emph{\bibinfo{journal}{Nature Phys.}} \textbf{\bibinfo{volume}{14}}, \bibinfo{pages}{739--744} (\bibinfo{year}{2018}).

\bibitem{deliancourt2019wavefront}
\bibinfo{author}{Deliancourt, E.} \emph{et~al.}
\newblock \bibinfo{title}{Wavefront shaping for optimized many-mode {Kerr} beam self-cleaning in graded-index multimode fiber}.
\newblock \emph{\bibinfo{journal}{Optics Express}} \textbf{\bibinfo{volume}{27}}, \bibinfo{pages}{17311--17321} (\bibinfo{year}{2019}).

\bibitem{LoveSnyder}
\bibinfo{author}{Snyder, A.~W.} \& \bibinfo{author}{Love, J.~D.}
\newblock \emph{\bibinfo{title}{Optical Waveguide theory}} (\bibinfo{publisher}{Chapman and Hall, London, UK}, \bibinfo{year}{1983}).

\bibitem{Zitelli}
\bibinfo{author}{Zitelli, M.}
\newblock \bibinfo{title}{A new thermodynamic approach to multimode fibre self-cleaning and soliton condensation}.
\newblock \emph{\bibinfo{journal}{arXiv}} \textbf{\bibinfo{volume}{2404}}, \bibinfo{pages}{00480} (\bibinfo{year}{2024}).

\bibitem{Crenn}
\bibinfo{author}{Crenn, J.~P.}
\newblock \bibinfo{title}{Optical study of the {${\rm EH}_{11}$} mode in a hollow circular oversized waveguide and {Gaussian} approximation of the far-field pattern}.
\newblock \emph{\bibinfo{journal}{Appl. Opt.}} \textbf{\bibinfo{volume}{23}}, \bibinfo{pages}{3428--3433} (\bibinfo{year}{1984}).

\bibitem{mangini}
\bibinfo{author}{Mangini, F.} \emph{et~al.}
\newblock \bibinfo{title}{Statistical mechanics of beam self-cleaning in grin multimode optical fibers}.
\newblock \emph{\bibinfo{journal}{Opt. Express}} \textbf{\bibinfo{volume}{30}}, \bibinfo{pages}{10850--10865} (\bibinfo{year}{2022}).
\newblock \urlprefix\url{https://opg.optica.org/oe/abstract.cfm?URI=oe-30-7-10850}.

\bibitem{Haus2}
\bibinfo{author}{Haus, H.~A.}, \bibinfo{author}{Fujimoto, J.~G.} \& \bibinfo{author}{Ippen, E.~P.}
\newblock \bibinfo{title}{Structures for additive pulse mode locking}.
\newblock \emph{\bibinfo{journal}{J. Opt. Soc. Am. B}} \textbf{\bibinfo{volume}{8}}, \bibinfo{pages}{2068--2076} (\bibinfo{year}{1991}).

\bibitem{Bespalov}
\bibinfo{author}{Bespalov, V.~I.} \& \bibinfo{author}{Talanov, V.~I.}
\newblock \bibinfo{title}{Filamentary structure of light beams in nonlinear liquids}.
\newblock \emph{\bibinfo{journal}{JETP Lett.}} \textbf{\bibinfo{volume}{3}}, \bibinfo{pages}{307--310} (\bibinfo{year}{1966}).

\bibitem{Malomed}
\bibinfo{author}{Sakaguchi, H.} \& \bibinfo{author}{Malomed, B.~A.}
\newblock \bibinfo{title}{Two-dimensional solitons in the {Gross-Pitaevskii} equation with spatially modulated nonlinearity}.
\newblock \emph{\bibinfo{journal}{Phys. Rev. E}} \textbf{\bibinfo{volume}{73}}, \bibinfo{pages}{026601} (\bibinfo{year}{2006}).

\end{thebibliography}
\newpage

\end{document}

\title{Bose-Einstein condensation of an optical thermodynamic system into a solitonic state\\ \\ Supplementary Materials}

\author{Jiaxuan Zhang}%
\thanks{These two authors contributed equally}
\author{Jintao Fan}%
\thanks{These two authors contributed equally}
\affiliation{Ultrafast Laser Laboratory, Key Laboratory of Opto-electronic Information Science and Technology of Ministry of Education, School of Precision Instruments and Opto-electronics Engineering, Tianjin University, 300072 Tianjin, China}%
\author{Chao Mei}
\affiliation{Department of Physics, School of Physical Science and Technology, Ningbo University, 315211, Zhejiang, China}%
\author{G{\"u}nter Steinmeyer}
\email{steinmey@mbi-berlin.de}
\affiliation{Max Born Institute for Nonlinear Optics and Short Pulse Spectroscopy, 12489 Berlin, Germany}%
\affiliation{Institut f{\"u}r Physik, Humboldt Universität zu Berlin, 12489 Berlin, Germany}%
\author{Minglie Hu}
\email{huminglie@tju.edu.cn}
\affiliation{Ultrafast Laser Laboratory, Key Laboratory of Opto-electronic Information Science and Technology of Ministry of Education, School of Precision Instruments and Opto-electronics Engineering, Tianjin University, 300072 Tianjin, China}%
\date{\today}

\section{Detailed evolution of beam profiles in the condensation process}

\begin{figure}[bt]
\centering
\includegraphics[scale=0.63, angle=270]{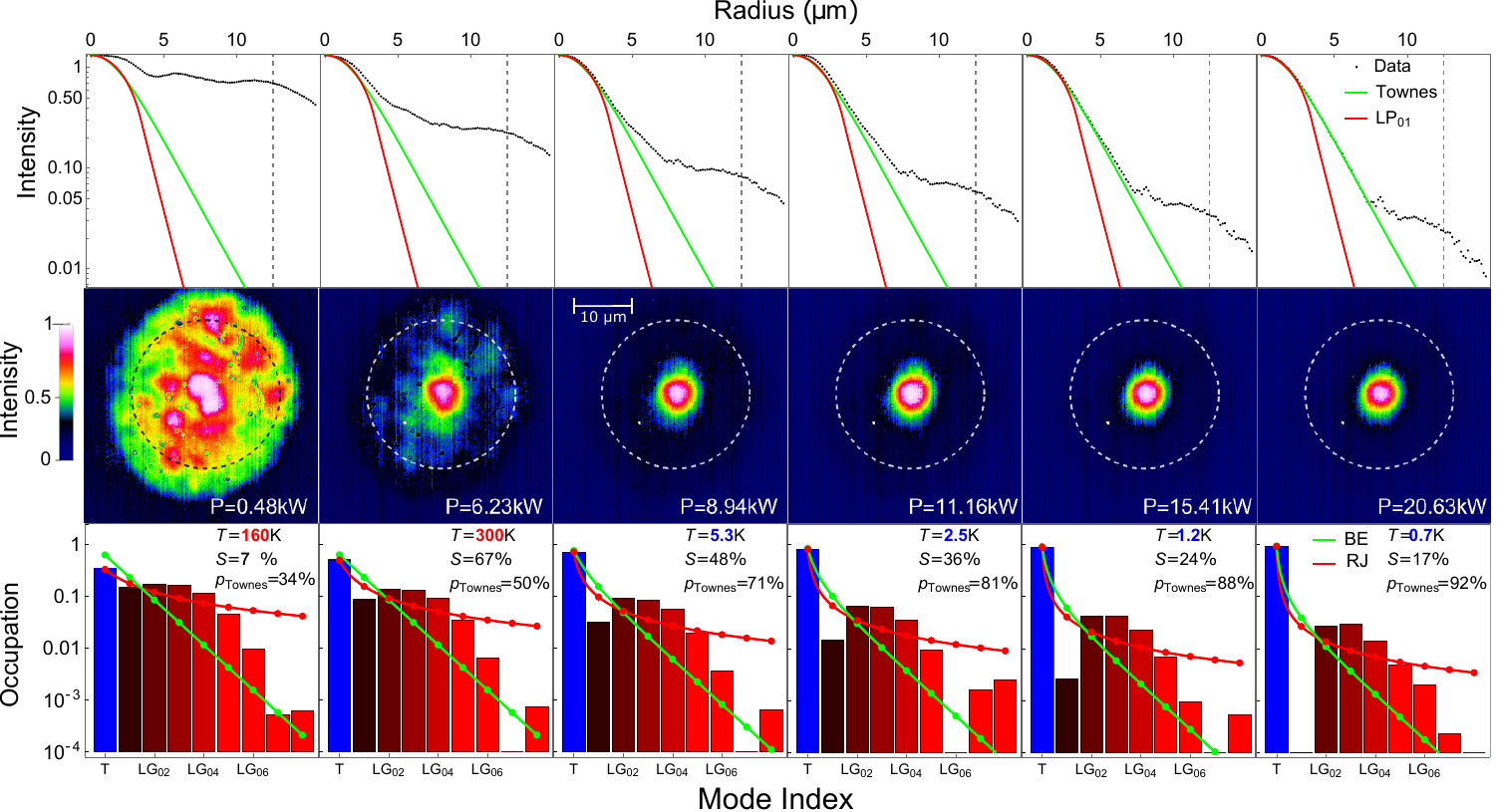}
\caption{Mode decomposition analysis of the dual-core fiber with dissipative core. {\bf Left column:} Decomposition into Townes mode and higher-order Gauss-Laguerre residuals ${\rm LG}_{01}$ -- ${\rm LG}_{08}$ on a linear scale and a log scale (insets). Green curves indicate a fitted Bose-Einstein distribution; red curves a Rayleigh-Jeans fit. Temperatures $T$ are shown as extracted from the chemical potential $\mu/k_{\rm B}$ of the Bose-Einstein fits. $S$ indicates Gibbs entropy relative to the maximum possible value of ${\rm ln}\,9$ and $p_{\rm Townes}$ the occupation of the Townes mode. {\bf Middle column:} original beam profile measurements. The core radius of 12.5 $\mu$m is indicated as a dashed circle in the beam profiles, and the peak power $P$ is indicated outside this circle. {\bf Right column:} azimuthally averaged near-field beam profiles extracted from beam profile measurements, respectively. Different functions have been fitted to measured radial beam profiles, as indicated by red and green curves.} 
\label{fig:S0}
\end{figure}

A more elaborate version of the decomposition analysis of the dual-core fiber with doped core is shown in Fig.~\ref{fig:S0}.

\section{Alternative Entropy Definitions}

In the main text, we use the Gibbs entropy definition \cite{entropy}
\begin{equation}
 S_{\rm Gibbs}= - k_B \sum_{i=1}^m p_i \log p_i,  \label{eq:Gibbs}
\end{equation}
where $0\le p_i \le 1$ is the relative occupation of the $i$th mode out of $n$, and the Boltzmann constant $k_B$ has been set to unity. Given that there is a total of $m$ modes, $S_{\rm Gibbs}$ may assume a maximum value of $\log m$ if all modes are equally populated, and all $p_i=1/m$. One can therefore define a relative entropy
\begin{equation}
S_{\rm rel}= - \frac{1}{\log m} \sum_{i=1}^m p_i \log p_i \in \left[0,1\right] 
\end{equation}
as was employed in Figs.~3 and 4 of the main text.

As an alternative definition, the use of Boltzmann entropy 
\begin{equation}
S_{\rm Boltzmann} = \log \prod_{i=1}^m \frac{(n_i+g_i-1)!}{n_i!(g_i-1)!} \label{eq:Boltzmann}
\end{equation}
has been suggested \cite{Zitelli}. Here the $n_i$ represents the total number of photons in the $i$th mode, and $g_i=2 i$ is the respective degeneracy factor. Moreover, $n_i$ is related to the total photon number $N$ by the relation 
\begin{equation}
N = \sum_{i=1}^m n_i  = \sum_{i=1}^m p_i \frac{\hat{P} \tau}{h \nu},  
\end{equation}
where $\hat{P}$ and $\tau$ are the pulse peak power and duration, respectively; $h$ is Planck's constant and $\nu$ the center wavelength. 

Given that the total number of photons 
\begin{equation}
N=\frac{\hat{P} \tau}{h \nu}
\end{equation}
is not a constant during our experiments, it appears reasonable to normalize Boltzmann's entropy to some constant number of photons, e.g., the number of photons at lowest energy $n_1$, yielding
\begin{equation}
S_{\rm B,norm} = \log \prod_{i=1}^m \frac{(p_i n_1+g_i-1)!}{(p_i n_1)!(g_i-1)!}. \label{eq:norm}
\end{equation}
The latter redefinition ensures constancy of the particle number and therefore the assumption of a closed thermodynamic system. Using Stirling's formula and assuming large photon numbers $n_i\gg g_i$, Eq.~(\ref{eq:Boltzmann})  can be approximated by
\begin{equation}
S_{\rm Boltzmann}  \approx  \Gamma_m + \sum_{i=1}^m (g_i-1) \log{n_i}  \\
\end{equation}
with a constant
\begin{equation}
\Gamma_m = \sum_{i=1}^m (g_i-1) [1-\log (g_i-1)].
\end{equation}
For the 9-mode decompositions in the main part of this paper, we compute $\Gamma_9\approx -112$. As this constant does not depend on the $n_i$ but only on the number of modes $n$ in the decomposition, it is customary to drop this in a comparative analysis \cite{Picozzi, Zitelli, wright2022physics}, yielding the following approximate version
\begin{equation}
S_{\rm Boltzmann}\approx \sum_{i=1}^m (g_i-1) \log n_i,
\end{equation}
or a normalized version
\begin{equation}
S_{\rm B,norm}\approx \sum_{i=1}^m (g_i-1) \log p_i, \label{eqn:Picozzi}
\end{equation} 
where another constant offset has been removed. While the original Boltzmann definition is robust against decompositions including zero occupations $n_i=0$, it should be noted that the approximate version in Eq.~(\ref{eqn:Picozzi}) diverges for $p_i=0$. Therefore, its usage for the analysis of experimental data is strongly discouraged. In the following, we therefore exclusively use Eqs.~(\ref{eq:Gibbs}, \ref{eq:Boltzmann}, and \ref{eq:norm}) for analysis of our experimental data.
\begin{figure}[tbh]
\centering
\includegraphics[width=\linewidth]{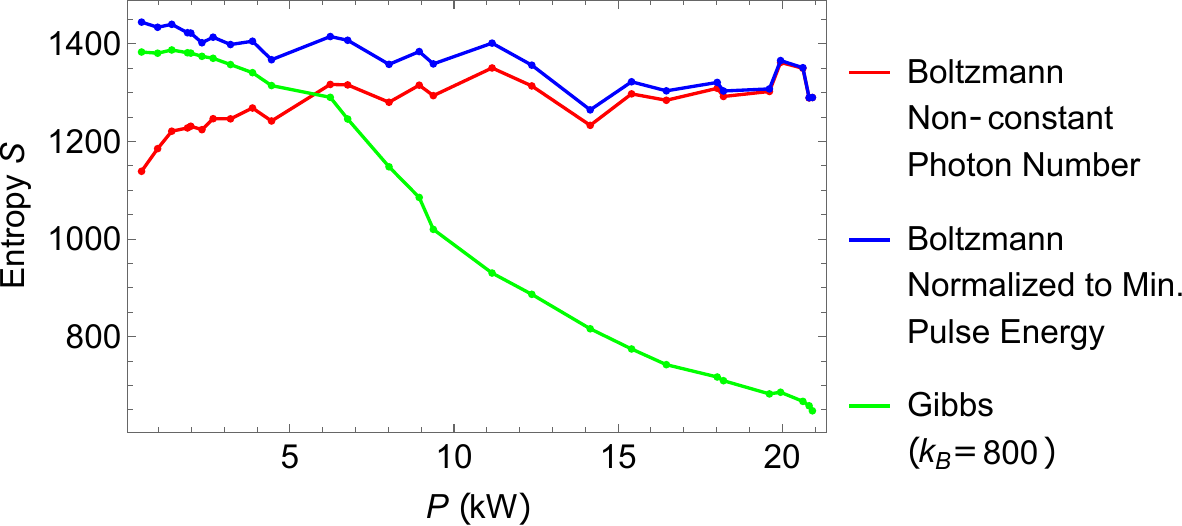}
\caption{Entropy vs. peak power for dissipative dual-core fiber. $k_B$ has been adjusted for display purposes, allowing immediate comparison with Boltzmann definitions.} 
\label{fig:S1}
\end{figure}

\begin{figure}[tbh]
\centering
\includegraphics[width=\linewidth]{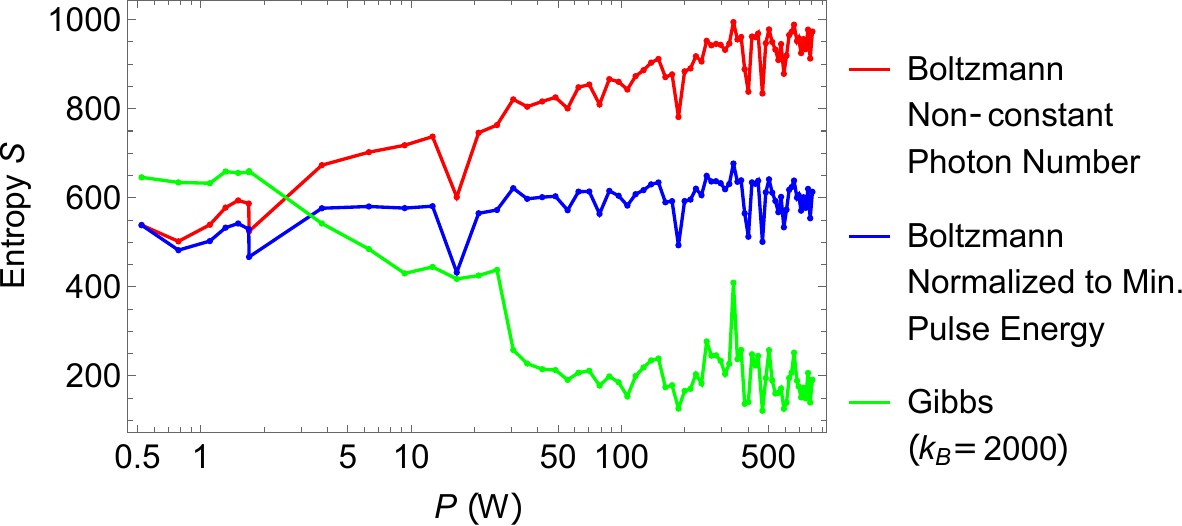}
\caption{Entropy vs. peak power for conservative dual-core fiber.  $k_B$ has been adjusted for display purposes, allowing immediate comparison with Boltzmann definitions.} 
\label{fig:S2}
\end{figure}

Figure \ref{fig:S1} and \ref{fig:S2} show the development of various entropy definitions for the dissipative fiber, respectively, according to the three selected definitions mentioned above. Here Gibbs entropy has been rescaled to the range of the Boltzmann definitions by choosing a somewhat arbitrary definition of $k_B$. Let us again repeat that such rescaling only causes a constant offset, which, however, does not affect the functional behavior of entropy with increasing power. 

Inspecting Figs.~\ref{fig:S1} and \ref{fig:S2} carefully, it immediately becomes apparent that the original Boltzmann entropy definition [Eq.~(\ref{eq:Boltzmann})] always increases with increasing power whereas the opposite behavior is observed for Gibbs entropy [Eq.(\ref{eq:Gibbs}]. As the increase of Boltzmann entropy may be artificially introduced by non-constancy of the photon number in the experiments, we rather use the normalized entropy [Eq.~(\ref{eq:norm})] to judge the significance of the entropy drop for either fiber. As the normalized entropy remains rather constant for the undoped fiber [Fig.~ \ref{fig:S2}], we conclude that the observed beam self-cleaning process at low powers is rather a thermalization process than a phase transition. We can therefore not conclude on any violation of the second law, and the thermodynamic system can be considered closed in good approximation. The drop in Gibbs entropy may therefore be explained by the rather small number of partitions in our modal decomposition.

The situation is clearly different for the dissipative fiber, where both Gibbs and normalized Boltzmann entropy decrease, Nevertheless, this effect is not surprising as absorptive effects are expected to reduce the number of photons, i.e., the thermodynamic system cannot be considered closed anymore. The increase of Boltzmann entropy therefore appears artificial.

\section{Alternative temperature evaluation}

\begin{figure}[tbh]
\centering
\includegraphics[width=\linewidth]{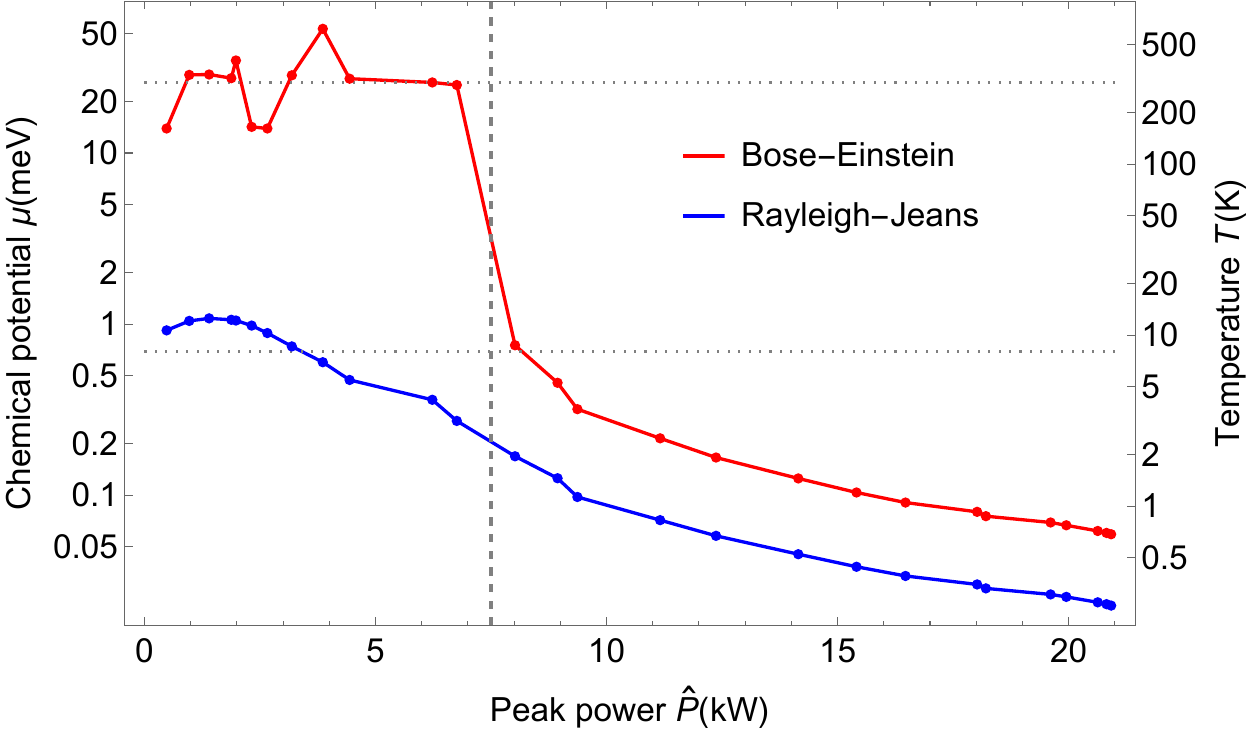}
\caption{Temperature evaluation from chemical potential $\mu$ of the dissipative fiber estimated from a fit of the modal distribution to Bose-Einstein (red) and Rayleigh-Jeans statistics (blue). Only the former exhibits a phase transition (dashed line) with a sudden drop in temperature by a factor ~50, as indicated by the dotted lines.} 
\label{fig:Stemp}
\end{figure}

In the main text, temperatures have always been estimated from the chemical potential $\mu$ in a fit to the Bose-Einstein distribution. Chemical potential can be converted to a temperature via $\mu=k_B T$, with the Boltzmann constant $k_B=8.6\times 10^{-5} {\rm eV}/{\rm K}$. Figure \ref{fig:Stemp} shows a comparison of temperatures extracted from Bose-Einstein and Rayleigh-Jeans fits to the measured modal distributions. While Rayleigh-Jeans indicates a smooth decrease of temperatures with increasing peak power $\hat{P}$, Bose-Einstein exhibits a sharp discontinuity at $\hat{P}\approx 7.5\,$kW, indicative of a phase transition. Above this threshold, both fits qualitatively yield identical results, yet, with a Rayleigh-Jeans temperature that is about 4 times lower than that extracted from Bose-Einstein statistics \cite{Zitelli}. 

\begin{figure}[tbh]
\centering
\includegraphics[width=0.7 \linewidth]{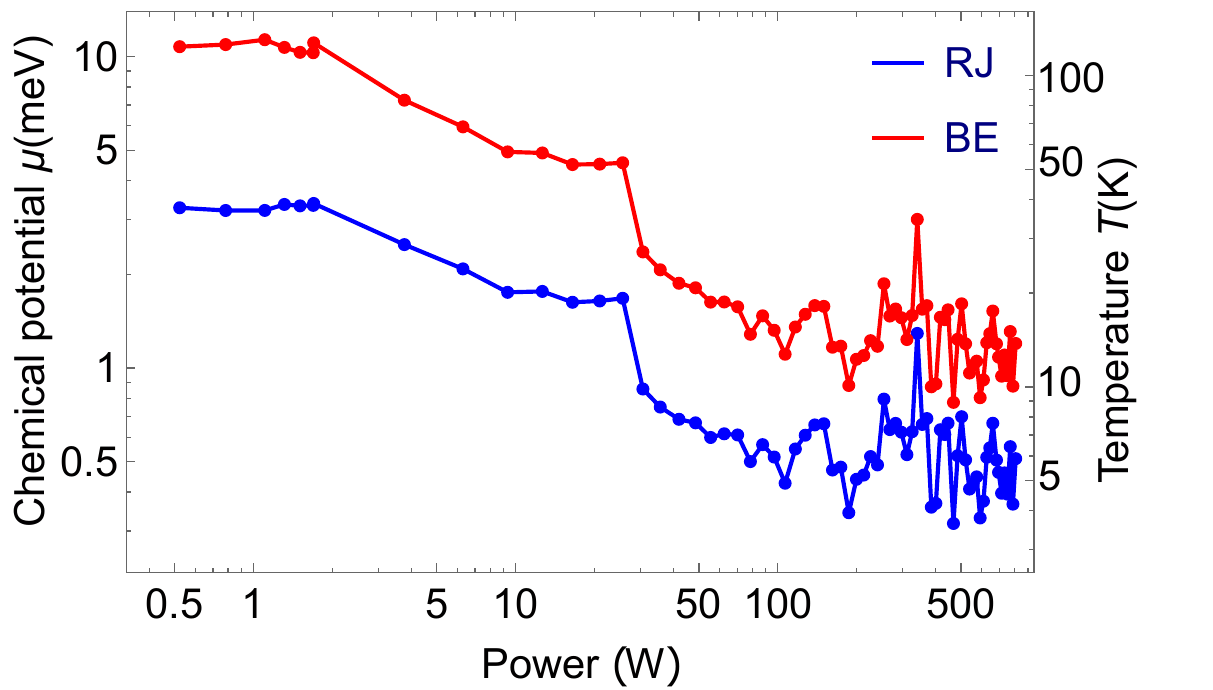}
\caption{Temperature evaluation from chemical potential $\mu$ of the conservative fiber estimated from a fit of the modal distribution to Bose-Einstein (red) and Rayleigh-Jeans statistics (blue).} 
\label{fig:Stemp1}
\end{figure}

\section{Numerical simulations}
To better understand the self-cleaning behavior of optical beams, we employ the following of the radial Gross-Pitaevskii equation (2D+1 nonlinear Schrödinger equation) to qualitatively describe the propagation of optical pulses in doped multimode optical fibers \cite{krupa2017spatial}. 
\begin{equation}
\frac{\partial A}{\partial{z}}-i\frac{1}{2k_0n_0}\nabla_{\bot}^2 A+i\left[\Delta \beta+k_0\bar{n}(\omega,r)\right]A=ik_0 n_2 A\left[(1-f_R)|A|^2+f_R\int_{-\infty}^t h(\tau)|A(t-\tau)|^2\text{d}\tau\right], \label{eq:NLSE}
\end{equation}
\begin{figure}[tbh]
\centering
\includegraphics[width=0.7 \linewidth]{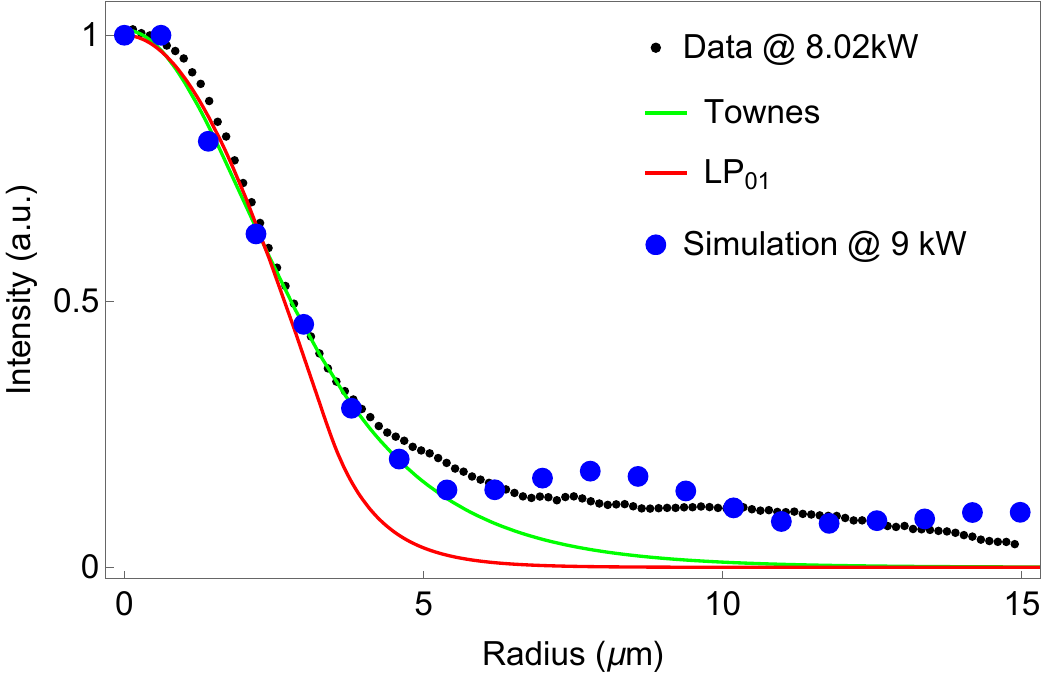}
\caption{Comparison of a numerical simulation of the radial Gross-Pitaevskii Equation with measured data.}
\label{fig:SIsimu}
\end{figure}
where $A(r,t,z)$ is the complex electric amplitude under slow-varying envelop approximation, $k_0=2\pi/\lambda_0$ is the wave number, $\lambda_0=1560$ nm the pump wavelength, $n_0$ the refractive index of inner core at pump wavelength. $\Delta \beta(\omega,r)=\beta(\omega,r)-\beta_0(\omega_0,r)-\beta_1(\omega_0,r)(\omega-\omega_0)$ with $\beta (\omega,r) = 2\pi n(\omega,r)/\lambda$ is the propagation constant of fundamental mode, where  $n(\omega,r) = \bar{n}(\omega,r)+ik_\text{e}(r)/2$ is the radially dependent complex refractive index and $\bar{n}$ is the real part of $n$ and $k_\text{e}$ is the extinction coefficient related to the linear loss of $\alpha$ by $\alpha=k_0k_\text{e}$. We further assume that $\alpha$ is frequency-independent and always equal to the value at the pump wavelength. For the inner and outer core $\alpha= 0.2$ and $\alpha=0.002$ dB/m were assumed, respectively. $\beta_0$ is the propagation constant at pump angular frequency $\omega_0$ after Taylor expansion of $\beta$. $\beta_1 (\omega_0,r)$ is the first-order derivative of $\beta$ at the pumping frequency. The nonlinear refractive index of doped silica fiber is $n_2 = 2\times 10^{-16}$ cm$^2$/W, $f_R = 0.18$ is the Raman fraction, and $h(\tau)$ is the Raman response function. The second, third, and fourth terms in Eq.~\ref{eq:NLSE} denote the linear effects including diffraction, dispersion, and potential imposed by the step-index profile. In contrast, the first and second terms at the right-hand side of Eq.~\ref{eq:NLSE} describe the nonlinear effects including the instantaneous Kerr and non-instantaneous Raman nonlinearities. In the simulations, we set the input spatial field to consist of 9 circularly symmetric linearly polarized modes, denoted as LP$_{0n}$, where $n$ ranges from 1 to 9. The degenerate modes within each LP mode are neglected. These 9 LP modes are set to be Gaussian-shaped in the time domain, with initial peak power ratios of 1/18, 1/18, 1/18, 1/18, 2/18, 3/18, 3/18, 3/18, and 3/18, respectively. When the peak power of input pulse is set to 9 kW, we observe good agreement between the simulated spatial distribution and experimental results obtained at an input power of 8 kW, as shown in Fig.~\ref{fig:SIsimu}. However, in the low-intensity regions of spatial profile, we notice an oscillatory structure in the simulation results. This finding deviates from the experimentally observed flat pedestal structure. This discrepancy arises probably because of the different input beam shapes. In the simulation, the input beam is completely a circularly symmetric beam combined from multiple Bessel functions, and its tail may not monotonically decrease. In contrast, in the experiments, the modes coupled into the fiber are comprised of substantially different shapes, exhibiting only slow changes with radius. We also observed that the potential term plays a crucial role in promoting the BSC, consistent with previous reports. However, apart from these deviations in the pedestal structure, the central part of the simulated beam profile shows excellent agreement with the experimentally found Townes profile for the dissipative fiber. This agreement between simulation, eigenmode solution of the GPE and experiments further strengthens the conclusions drawn in the main text of this article.
\bibliographystyle{naturemag}
\bibliography{Ref}